\documentclass[twocolumn,showpacs,preprintnumbers,bibnotes]{revtex4}
\usepackage{graphicx}
\usepackage{dcolumn}
\usepackage{bm}
\usepackage{amssymb}
\usepackage{amsmath}
\usepackage{epsfig}
\usepackage{bm}

\begin{document}


\title{Characterizing Temporal Distinguishability of an N-Photon State\\
 by Generalized Photon Bunching Effect with Multi-Photon Interference}

\author{Z. Y. Ou}\email{zou@iupui.edu}
 \affiliation{Department of Physics,Indiana
University-Purdue University Indianapolis \\ 402 N. Blackford
Street, Indianapolis, IN 46202}

\date{\today}

\begin{abstract}
The complementary principle of quantum mechanics relates
qualitatively the visibility of quantum interference with path
indistinguishability. Here we propose a scheme of constructive
quantum interference involving superposition between an $N$-photon
state and a single-photon state to characterize quantitatively the
degree of temporal distinguishability of the $N$-photon state.
This scheme is based on a generalized photon bunching effect. Such
a scheme can be extended to other more general cases.

\end{abstract}

\maketitle

\section{Introduction}

The complementary principle of quantum mechanics was first
proposed by Bohr \cite{bo} to deal with the wave-particle duality
of quantum particles. On the one hand, it successfully explained
the peculiar quantum behavior of particles in interference. On the
other hand, it only provides a qualitative description of quantum
interference process. The problem stems from the lack of a
quantitative definition of distinguishability. Efforts were made
to find such a definition with some success \cite{scu,un,rem,zei}.

The above mentioned discussions of the complementary principle
were mostly confined in fundamental conceptual study and in
interference involving only one particle. However, recent
interests on quantum information involve quantum interference of
multiple particles \cite{bou}, especially in the context of linear
optical quantum computing with qubits realized by
photons\cite{klm}. An issue thus arises about distinguishability
among the photons that may degrade the quantum interference
effects, leading to poor performance of the quantum operations. So
it is desirable to study photon distinguishability qualitatively
and to find its relation with multi-photon quantum interference
effect.

The first investigation of the effect of photon distinguishability
on multi-photon interference was performed by Grice and Walmsley
\cite{gri} with an analysis on a two-photon polarization
Hong-Ou-Mandel interferometer \cite{hom}. A more complicated
four-photon case was studied by Ou et al \cite{rhe1,rhe2} and
later by Tsujino et al \cite{tsu,ou2} with concerns about the
distinguishability between two pairs of photons.

Recently, the current author \cite{ou} made an attempt to
generalize to the above discussion to a system of arbitrary number
of photons. A degree of temporal distinguishability is
quantitatively defined and  a destructive multi-photon
interference method is proposed that relies on a quantum state
projection measurement \cite{sun1,sun2,res} to experimentally
measure it. Subsequent experimental demonstrations \cite{xia,liu}
confirmed some of the predictions. It was shown \cite{ou} that the
visibility of interference is proportional the number of
indistinguishable photons in a simple situation. But since the
visibility is bounded by 1, more accurate measurement on the
visibility is required to distinct various scenarios of different
photon distributions, especially when the photon number is large.
The accuracy problem is compounded by the fact that destructive
interference in this scheme makes the measured quantity small, so
that it requires long recording time for good accuracy.
Furthermore, the scheme of quantum state projection measurement is
complicated in structure and requires phase shifters with precise
values. Another scheme was recently discussed in Ref.\cite{ou07}
that relies on a generalized Hong-Ou-Mandel interference effect
with asymmetric beam splitter. This scheme needs less optical
elements and thus significantly simplifies the optical
arrangement. But since this new scheme is based on destructive
interference, it still suffers the problem of low count rate at
maximum interference effect and thus low measurement accuracy.
Another disadvantage in this new scheme is that we need to control
the precise value of the transmissivity of the beam splitter,
which depends on the total photon number, in order to achieve
complete destructive multi-photon interference.

It was demonstrated that a photon pair bunching effect can be used
to characterize temporal distinguishability between two pairs of
photons \cite{rhe1,rhe2}.  In this case, constructive four-photon
interference for two pairs of photons leads to five-fold increase
in four-photon coincidence when the two pairs are
indistinguishable, whereas the state of two separated pairs
produces only three-fold increase. The enhancement factor is
expected to be bigger for larger photon number due to Bose
statistics. Largeness of the measured quantity leads to a good
accuracy in the four-photon coincidence measurement.

In this paper, we generalize the study of the photon bunching
effect to arbitrary photon number. We find that the enhancement
factor in photon bunching is due to constructive interference and
can be used to characterize the temporal distinguishability of
photons. The enhancement factor is not sensitive to the
experimental parameters and the optical arrangement is relatively
simple. This scheme seems to overcome all the shortcomings of
previous schemes. The paper is arranged as follows: In Sect.II, we
will discuss stimulated emission as a photon bunching effect due
to constructive interference and exploit it for characterizing
temporal distinguishability of incoming photons. We will also
discuss its analogy with a beam splitter. This is a simple single
mode analysis. In Sect.III, we will perform the more rigorous
multi-mode analysis and confirm the results from the simple single
mode analysis. In Sect.IV, we will consider other scenarios of
photon temporal distributions and derive the corresponding
enhancement factor. We end the paper with a discussion.

\section{Stimulated emission as a multi-photon constructive
interference effect}

Recently, it was pointed out \cite{sun07} that stimulated emission
can be interpreted as a result of multi-photon constructive
interference: when $N$ input photons are indistinguishable from
the photon emitted by the excited atom, constructive interference
leads to a factor of $N$ enhancement in the atomic emission rate
from spontaneous emission. The enhanced emission is due to
stimulated emission. On the other hand, if the input photons are
completely distinguishable from the photon emitted by the atom, no
enhancement occurs and the atom makes only spontaneous emission.

If the input photons are partially indistinguishable from the
emitted photon, only the indistinguishable part will give rise to
the stimulated emission. Therefore, such a scheme can be used to
characterize quantitatively the degree of distinguishability of
the input photons. To see how this works, we consider an excited
atom modelled as a phase insensitive quantum amplifier with small
gain \cite{cav}:
\begin{eqnarray}
\hat a_s^{(out)} = G\hat a_s +g \hat a_0^{\dag}, \label{1}
\end{eqnarray}
where $\hat a_0$ represents all the internal modes of the
amplifier and it is usually independent of the signal mode $\hat
a_s$ and is in vacuum. To preserve the commutation relation, we
need $|G|^2 - |g|^2 =1$ and for small gain, $|g|<<1$. The related
evolution operator for the system has the form of
\begin{eqnarray}
\hat U = \exp\{\eta \hat a_s^{\dag}\hat a_0^{\dag}-h.c.\} \approx
1 + (g \hat a_s^{\dag}\hat a_0^{\dag} + h.c.)\label{2}
\end{eqnarray}
with $g\approx \eta$.

With a vacuum input of $|0\rangle$, we have the output state
\begin{eqnarray}
|\Phi\rangle_{out}^{(0)} =\hat U |0\rangle  \approx |0\rangle + g
|1\rangle_s\otimes|1\rangle_0.\label{3}
\end{eqnarray}
This gives the spontaneous emission probability of $|g|^2$. When
the input is an $N$-photon state $|N\rangle_s\otimes|0\rangle_0$,
we have
\begin{eqnarray}
|\Phi\rangle_{out}^{(1)} &\approx & |N\rangle_s|0\rangle_0 + g
(\hat a_s^{\dag}|N\rangle_s)\otimes(\hat a_0^{\dag}|0\rangle_0
)\cr &= &|N\rangle_s|0\rangle_0 + g\sqrt{N+1}
|N+1\rangle_s\otimes|1\rangle_0.\label{4}
\end{eqnarray}
The probability becomes $(N+1)|g|^2$. The stimulated emission
helps to enhance the emission rate by a factor of $N+1$.

In Eq.(\ref{4}), the input $N$ photons are all in the same mode as
the mode $\hat a_s$ of the amplifier. However, if some of the
input photons are in different modes from the mode $\hat a_s$  of
the amplifier, these photons are not coupled to the amplifier and
cannot stimulate the emission of the amplifier. Mathematically, we
have the input as $|m\rangle_s|N-m\rangle_{s'}|0\rangle_0$ and the
output state as
\begin{eqnarray}
|\Phi\rangle_{out}^{(1)'} &\approx &
|m\rangle_s|N-m\rangle_{s'}|0\rangle_0 \cr &&\hskip 0.2in+ g (\hat
a_s^{\dag}|m\rangle_s)\otimes(|N-m\rangle_{s'})\otimes(\hat
a_0^{\dag}|0\rangle_0 )\cr &=
&|m\rangle_s|N-m\rangle{s'}|0\rangle_0 \cr &&\hskip 0.2in+
g\sqrt{m+1} |m+1\rangle_s|N-m\rangle_{s'}|1\rangle_0.\label{5}
\end{eqnarray}
The enhancement factor is now $m+1$. In the special cases when
$m=0, N$, we recover Eqs.(\ref{3}, \ref{4}), respectively.
Therefore, spontaneous emission corresponds to the case when the
input photons are completely distinguishable from the photon
emitted from the amplifier whereas stimulated emission occurs when
the input photons are indistinguishable from the photon emitted by
the amplifier.

Notice that the enhancement factor $m+1$ is linearly related to
the number of indistinguishable photons. Thus by observing the
size of the enhancement, we can quantitatively characterize the
degree of distinguishability. However, the mode of the amplifier
is somewhat complicated, which makes this scheme hard to
implement.

\begin{figure}[htb]
\begin{center}
\includegraphics[width= 3in]{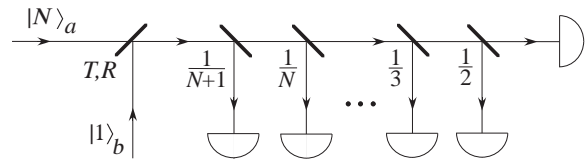}
\end{center}
\caption{ Generalized photon bunching effect for characterizing
the distinguishability of photons.} \label{fig1}
\end{figure}

We can circumvent this problem with linear optics. As discussed
before, the enhancement effect in stimulated emission is due to
photon indistinguishability and is the result of constructive
multi-photon interference, which can then be mimicked by a
lossless beam splitter, as shown in Fig.1. Ref.\cite{sun07} showed
that the result in the scheme of Fig.1 for $N+1$ indistinguishable
photons is the same as the stimulated emission process described
by Eq.(\ref{4}) with an enhancement factor of $N+1$ compared to
the situation when the $N$ photons are distinguishable from the
single photon at the other side of the beam splitter.

For the case when $m\ne N$, i.e., the case when some of the input
$N$ photons are distinguishable, we may use a similar input state
as in Eq.(\ref{5}), i.e., $|N-m\rangle_{a'}\otimes
|m\rangle_a|1\rangle_b$. In this state, the $N-m$ photons are
distinguishable from the $m$ photons and the single photon from
the other side.  Since the state $|m\rangle_a|1\rangle_b$ is the
same as the result of stimulated emission with input state of
$|m\rangle$ and the state $|N-m\rangle_{a'}$ has no enhancement
effect, the overall enhancement factor is simply $m+1$, exactly
the same as the case of stimulated emission given in Eq.(\ref{5}).

The enhancement effect with a beam splitter in Fig.1 is a
generalized photon bunching effect and it can be similarly used
for the characterizing the degree of photon distinguishability.

However, the above analysis is a single-mode analysis. To fully
prove its validity, we need a multi-mode analysis.

\section{Multi-mode analysis of the multi-photon bunching effect}

In this multi-mode analysis, we will only concentrate on the
temporal/spectral mode and ignore other modes such as spatial and
polarization modes. However, the generalization is
straightforward.

Let us consider the scheme in Fig.1 where an  $N$-photon state and
a single-photon state  enter a beam splitter from two separate
sides (labelled as $a$ and $b$). According to Ref.\cite{ou}, an
arbitrary $N+1$-photon state of a wide spectral range can be
expressed for multi-mode analysis as
\begin{eqnarray}
&&|\Phi_{N,1}\rangle = {\cal N}^{-1/2}\int d\omega_0
d\omega_1d\omega_2...d\omega_N \Phi(\omega_0;\omega_1, ...,
\omega_N)\cr &&\hskip 0.8in \times\hat
b^{\dag}(\omega_0)a^{\dag}(\omega_1)\hat a^{\dag}(\omega_2)...\hat
a^{\dag}(\omega_N)|0\rangle,~~~\label{Phi-state}
\end{eqnarray}
where the normalization factor ${\cal N}$ is given by
\begin{eqnarray}
&&{\cal N} = \int d\omega_0d\omega_1...d\omega_N
\Phi^*(\omega_0;\omega_1, ..., \omega_N)\times\cr &&\hskip
1.0in\times\sum_{\mathbb{P}} \Phi(\omega_0; \mathbb{P}\{\omega_1,
..., \omega_N\})\label{N}
\end{eqnarray}
with $\mathbb{P}$ as the permutation operator on the indices of 1,
2, ..., $N$. The sum is over all possible permutations. $\hat a$
and $\hat b$ represent the input modes $a$ and $b$ of the beam
splitter, respectively.

Let us consider the situation when the single photon from input
port $b$ overlaps temporally with $m$ photons from the $N$ input
photons at input port $a$ and the rest of the $(N-m)$ photons in
side $a$ are completely distinguishable in time from the $(m+1)$
photons. From Ref.\cite{ou}, we learn that the $N+1$-photon wave
function satisfies the permutation symmetry relations:
\begin{eqnarray}
&&\Phi(\omega_0; \omega_1,..., \omega_{N}) \cr &&\hskip 0.4in=
\Phi(\mathbb{P}\{\omega_0; \omega_1,..., \omega_{m}\},
\omega_{m+1},..., \omega_{N}),~~~~~~\label{perm}
\end{eqnarray}
for all permutation operation $\mathbb{P}$ and the orthogonal
relations:
\begin{eqnarray}
&&\int d\omega_0d\omega_1...d\omega_N \Phi^*(\omega_0;
\omega_1,..., \omega_{N})  \cr &&\hskip 0.7in
\times\Phi(\mathbb{P}_{kj}\{\omega_0; \omega_1,...,
 \omega_{N}\}) = 0,~~~~~~~\label{orth}
\end{eqnarray}
where $\mathbb{P}_{kj}$ interchanges the indices $k,j$ with $k\le
m, j\ge m+1$. Here Eq.(\ref{perm}) is for indistinguishability
among the $m+1$ photons whereas Eq.(\ref{orth}) is for temporal
distinguishability between the $m+1$ photons and the remaining
$N-m$ photons. Eqs.(\ref{perm}, \ref{orth}) can be equivalently
written in time domain as
\begin{eqnarray}
&&G(t_0; t_1,..., t_{N}) \cr &&\hskip 0.4in= G(\mathbb{P}\{t_0;
t_1,..., t_{m}\}, t_{m+1},..., t_{N}) ~~~~~~\label{perm2}
\end{eqnarray}
and
\begin{eqnarray}
&&\int dt_0dt_1...dt_N G^*(t_0; t_1,..., t_{N}) \cr &&\hskip 0.7in
\times G(\mathbb{P}_{kj}\{t_0; t_1,...,
 t_{N}\}) = 0,~~~~~~~\label{orth2}
\end{eqnarray}
where the notations are the same as before and
\begin{widetext}
\begin{eqnarray}
G(t_0;t_1,...,t_N)\equiv {1\over (2\pi)^{(N+1)/2}}\int
d\omega_0...d\omega_N \Phi(\omega_0;\omega_1,...,\omega_N)
e^{-i(\omega_0t_0+\omega_1t_1+...+\omega_Nt_N)}.\label{G}
\end{eqnarray}

The $N+1$-photon coincidence rate of the $N+1$ detectors in Fig.1
is proportional to a time integral of the correlation function of
\cite{gla}
\begin{eqnarray}
\Gamma^{(N+1)}(t_0,t_1,...,t_{N})= \langle \Phi_{N,1}| \hat
E_1^{(o)\dag}(t_{N})...\hat E_{1}^{(o)\dag}(t_1)\hat
E_{1}^{(o)\dag}(t_0) \hat E_1^{(o)}(t_0)\hat E_1^{(o)}(t_1)...\hat
E_{1}^{(o)}(t_{N})|\Phi_{N,1}\rangle.~~~~~~\label{Gamma}
\end{eqnarray}
where
\begin{eqnarray}
\hat E_{1}^{(o)}(t)=\sqrt{T}\hat E_a(t)+\sqrt{R}\hat E_b(t)
~~~~~{\rm with}~~~~\hat E_c(t)=(1/\sqrt{2\pi})\int d\omega ~\hat
c(\omega) e^{-i\omega t}~~~~~~(c=a,b).\label{Eab}
\end{eqnarray}
Let us first evaluate $\hat E_1^{(o)}(t_0)\hat E_1^{(o)}(t_1) ...
\hat E_{1}^{(o)}(t_{N})|\Phi_{N,1}\rangle$, which has the form of
\begin{eqnarray}
\hat E_1^{(o)}(t_0)\hat E_1^{(o)}(t_1)...\hat
E_{1}^{(o)}(t_{N})|\Phi_{N,1}\rangle = T^{N/2}R^{1/2}\sum_{k=0}^N
\mathbb{P}_{0k}\{ \hat E_b(t_0) \hat E_a(t_1)...\hat
E_a(t_N)\}|\Phi_{N,1}\rangle\label{EoN}
\end{eqnarray}
It is straightforward to show that for the state in
Eq.(\ref{Phi-state}), we have
\begin{eqnarray}
 \hat E_b(t_0) \hat E_a(t_1)...\hat
E_a(t_N)|\Phi_{N,1}\rangle &=& {{\cal N}^{-1/2}\over
(2\pi)^{(N+1)/2}}\int d\omega_0...d\omega_N
\sum_{\mathbb{P}}\Phi(\omega_0;\mathbb{P}\{\omega_1,...,\omega_N\})
e^{-i(\omega_0t_0+...+\omega_Nt_N)}|0\rangle\cr &=& {\cal
N}^{-1/2} {\cal G}(t_0;t_1,...,t_N)|0\rangle,\label{EN}
\end{eqnarray}
where
\begin{eqnarray}
{\cal G}(t_0;t_1,...,t_N)\equiv
\sum_{\mathbb{P}}G(t_0;\mathbb{P}\{t_1,...,t_N\}).\label{GV}
\end{eqnarray}

The overall $(N+1)$-photon coincidence probability is then a time
integral of the $\Gamma$-function in Eq.(\ref{Gamma}):
\begin{eqnarray}
P_{N+1}=\int
dt_0dt_1...dt_N\Gamma^{(N+1)}(t_0,t_1,...,t_{N}).\label{PN}
\end{eqnarray}
With Eqs.(\ref{G}, \ref{EoN}, \ref{EN}, \ref{GV}), we obtain
\begin{eqnarray}
P_{N+1}=T^NR{\cal N}^{-1}\int
dt_0dt_1...dt_N\sum_{k,j}\mathbb{P}_{0k}\{{\cal
G}^*(t_0,t_1,...,t_{N})\}\mathbb{P}_{0j}\{{\cal
G}(t_0,t_1,...,t_{N})\}=T^NR\bigg(\sum_{k=j} + \sum_{k\ne
j}\bigg).\label{PN2}
\end{eqnarray}
It is straightforward to find that the first term in
Eq.(\ref{PN2}) is
\begin{eqnarray}
\sum_{k=j} = {\cal N}^{-1} (N+1) \int
dt_0dt_1...dt_N\Big|\mathbb{P}_{0k}\{{\cal
G}(t_0,t_1,...,t_{N})\}\Big|^2={\cal N}^{-1}(N+1)\int
dt_0dt_1...dt_N \Big|{\cal
G}(t_0,t_1,...,t_{N})\Big|^2,\label{first}
\end{eqnarray}
where we switched the integral variables $t_0$ and $t_k$. Using
Eq.(\ref{GV}) and the fact that $\sum_{\mathbb{P}}$ is over all
permutations in $\{t_1,...,t_N\}$, Eq.(\ref{first}) becomes
\begin{eqnarray}
\sum_{k=j} = {\cal N}^{-1}(N+1)\int dt_0dt_1...dt_N N!
G^*(t_0;t_1,...,t_{N})\sum_{\mathbb{P}}G(t_0;\mathbb{P}\{t_1,...,t_{N}\})=
{\cal N}^{-1}(N+1)N!{\cal N}= (N+1)!,\label{first2}
\end{eqnarray}
where for an arbitrary permutation in ${\cal G}^*$, we also
permute the variables of integral in the same way. Since ${\cal
G}$ is unchanged under any such permutation, we obtain $N!$
identical terms in Eq.(\ref{first2}).

Next, let us consider the second term in Eq.(\ref{PN2}). We
evaluate one arbitrary term in the sum ($k\ne j$):
\begin{eqnarray}
\int dt_0...dt_{N} \sum_\mathbb{P} G^*(t_k,\mathbb{P}\{t_1,
...,t_0 ,..., t_{N}\}) \sum_{\mathbb{P}^{\prime}}
G(t_j,\mathbb{P}^{\prime}\{t_1, ..., t_0, ..., t_{N}\}).
\end{eqnarray}
\end{widetext}
Because of the permutation properties in Eqs.(\ref{perm2},
\ref{orth2}) and $t_k\ne t_j$, the only way to obtain a non-zero
integral is for $t_k$ in $\mathbb{P}'\{t_1, ..., t_0,..., t_{N}\}$
to be permuted to the first $m$ positions by
$\mathbb{P}^{\prime}$. Then we can use the permutation relation in
Eq.(\ref{perm2}) to interchange it with $t_j$ so that for these
$\mathbb{P}^{\prime}$s, we have:
\begin{eqnarray}
&&G(t_j,\mathbb{P}^{\prime}\{t_1, ..., t_0 ,..., t_{N}\})=
G(t_k,\mathbb{P}^{\prime}\{t_1, ..., t_0, ..., t_{N}\}).\cr &&
\end{eqnarray}
Permutation  by $\mathbb{P}^{\prime}$ to other positions for $t_k$
cannot be interchanged with $t_j$ and by the orthogonal relation
in Eq.(\ref{orth2}), the integral is zero. Since this is only
about $t_k$ in $\mathbb{P}^{\prime}$, other $N-1$ time variables
in $\mathbb{P}^{\prime}$ are free to move. So, there will be
$m(N-1)!$ permutation terms in the sum over $\mathbb{P}^{\prime}$
that are nonzero and, as before in Eq.(\ref{first2}), they will
all have the same time integral of $(N+1){\cal N}$. Therefore, the
second term in Eq.(\ref{PN2}) is equal to
\begin{eqnarray}
 \sum_{k\ne j} = {\cal N}^{-1}  \sum_{k\ne j} m(N-1)!{\cal N} = m (N+1)!.\label{second}
\end{eqnarray}
Combining Eqs.(\ref{first2}, \ref{second}), we have
\begin{eqnarray}
P_{N+1} = T^NR (1+m) (N+1)!.\label{PN1}
\end{eqnarray}
The case of $m=0$ corresponds to the situation when the single
photon of port $b$ is completely distinguishable from all the $N$
photons from port $a$, which gives no interference and sets a
baseline for reference. Thus the enhancement factor for $m\ne 0$
is $m+1$, in agreement with the single mode analysis. Note that
the enhancement factor does not depend on the transmissivity $T$
but the detection probability $P_{N+1}$ does. From Eq.(\ref{PN1}),
we find the maximum detection probability at $T=N/(N+1)$.

For the intermediate case when there are some partial
indistinguishability among the photons, the wave function $G$ does
not satisfy Eqs.(\ref{perm2}, \ref{orth2}). We will not obtain a
simple close form as Eq.(\ref{PN1}). However, for a special case
when all the $N$ photons from port $a$ are indistinguishable from
each other and they are only partially indistinguishable from the
single photon input from port $b$, we have the permutation
symmetry relation:
\begin{eqnarray}
G(t_0; t_1,..., t_{N}) = G(t_0; \mathbb{P}\{t_1,...,
t_{N}\})~~~~~~\label{perm3}
\end{eqnarray}
but no orthogonal relation similar to Eq.(\ref{orth2}). Then
Eq.({\ref{GV}) becomes
\begin{eqnarray}
{\cal G}(t_0;t_1,...,t_N)= N! G(t_0;t_1,...,t_N).\label{GV2}
\end{eqnarray}
With some manipulations, we find Eq.(\ref{PN2}) becomes
\begin{eqnarray}
P_{N+1}&=&T^NR(N+1)!(1+N{\cal V}_N)\cr &=& P_{N+1}^{cl}(1+N{\cal
V}_N)\label{PN3}
\end{eqnarray}
with
\begin{eqnarray}
{\cal V}_N &\equiv &{\int dt_0dt_1...dt_N
G^*(t_0;t_1,...,t_{N})G(t_1;t_0,...,t_{N})\over \int
dt_0dt_1...dt_N |G(t_0;t_1,...,t_{N})|^2}\cr &=& {\int d\omega_0
...d\omega_N \Phi^*(\omega_0;\omega_1, ...,
\omega_N)\Phi(\omega_1;\omega_0, ..., \omega_N)\over \int
d\omega_0 ...d\omega_N |\Phi(\omega_0;\omega_1, ...,
\omega_N)|^2}.\cr &&\label{VN}
\end{eqnarray}
Thus the enhancement factor is
\begin{eqnarray}
{P_{N+1}\over P_{N+1}^{cl}}= 1+N{\cal V}_N.\label{Enh}
\end{eqnarray}
Note that when ${\cal V}_N=1$, we have the maximum enhancement
factor of $N+1$, indicating complete indistinguishability whereas
when ${\cal V}_N=0$, there is no enhancement effect, due to
complete distinguishability. So the quantity ${\cal V}_N$ gives
the degree of distinguishability between the single photon in port
$b$ and the $N$ photons in port $a$.

The more general case is when there are $m$ indistinguishable
photons among the $N$ photons in port $a$ and the $m$ photons have
partial indistinguishability from the single photon in port $b$.
To simplify the case, we further assume that the other $N-m$
photons from port $a$ are completely distinguishable from the
above $m+1$ photons. For this case, we can show similar to
Eq.(\ref{Enh}) that the enhancement factor is
\begin{eqnarray}
{P_{N+1}\over P_{N+1}^{cl}}= 1+m{\cal V}_m,\label{Enhm}
\end{eqnarray}
where ${\cal V}_m$ is defined in Eq.(\ref{VN}) but with the wave
function $G$ satisfying the permutation symmetry relation in
Eq.(\ref{perm2}) in stead of Eq.(\ref{perm3}).

Now the experimental procedure to measure the distinguishability
of the $N$-photon state is depicted in Fig.2, where we scan the
relative delay of the single photon in port $b$ with respect to
the $N$-photon state in port $a$. Whenever the single photon scans
through $m$ indistinguishable photons, the $N+1$ coincidence count
shows a bump of size $m$ relative to the baseline. In this way, we
can characterize the temporal distinguishability of the $N$-photon
state. The width of the bump is determined by the function ${\cal
V}_m$.

\begin{figure}[htb]
\begin{center}
\includegraphics[width= 3in]{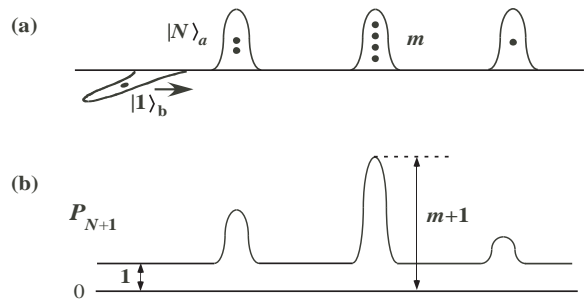}
\end{center}
\caption{(a) A temporal distribution with well-separated groups of
$N$ photons and (b) the corresponding normalized $P_{N+1}$ as the
position of the single photon is scanned.}
\end{figure}
\section{More general case of $|N_a, M_b\rangle$}

For the more general input state of $|N_a, M_b\rangle$, we can use
the analysis similar to Eqs.(\ref{4}, \ref{5}). First, if the
$N+M$ photons are indistinguishable, then from the quantum theory
of a lossless beam splitter \cite{cam}, we may find the
probability of finding all $N+M$ photons in one output side of the
beam splitter as
\begin{eqnarray}
P_{N+M} = T^{N}R^M(N+M)!/N!M!.\label{6}
\end{eqnarray}
But when the incoming $N$ photons are distinguishable from the $M$
photons, the $N+M$ photons act like classical particles and follow
the probability law. The corresponding classical probability is
then
\begin{eqnarray}
P_{N+M}^{cl} = T^{N}R^M.\label{7}
\end{eqnarray}
So the enhancement factor due to quantum interference is
\begin{eqnarray}
{P_{N+M}\over P_{N+M}^{cl}} = {(N+M)!\over N!M!}.\label{8}
\end{eqnarray}
A special case is for $N=M=2$, which gives the ratio of 6. This is
the photon pair bunching effect experimentally demonstrated by Ou
et al \cite{rhe1}.

The photon bunching enhancement factor in Eq.(\ref{8}) is for all
the photons involved to be indistinguishable. When some of the
photons are distinguishable, the enhancement factor will decrease.
The most general scenario is that some of the $N$ photons at input
port $a$ are indistinguishable from some of the $M$ photons at
side $b$. Let's break the $N$ photons and the $M$ photons into
$k+1$ groups, respectively, namely, $N=n_1+...+n_k+n_{k+1}$ and
$M=m_1+...+m_k+m_{k+1}$. In these groups, the $n_i$ photons are
indistinguishable from $m_i$ photons with $i=1,2, ..., k$ and
$\{n_i,m_i\}$ group of photons are distinguishable from
$\{n_j,m_j\}$ group of photons with $i\ne j$. Furthermore, the
$n_{k+1}$ photons are distinguishable from $m_{k+1}$ photons. Such
a situation is depicted in Fig.3.

\begin{figure}[htb]
\begin{center}
\includegraphics[width= 3in]{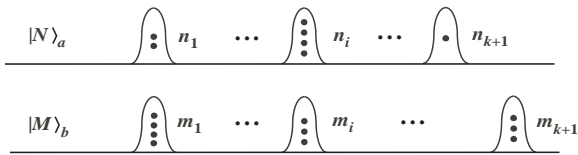}
\end{center}
\caption{Temporal distributions for photons from input sides $a$
and $b$, respectively.}
\end{figure}

In analogy to the case of stimulated emission described in
Eq.(\ref{5}), we may write the input state to the beam splitter as
\begin{eqnarray}
&&|\Phi\rangle_{in} =
|n_{k+1}\rangle_a^{(k+1)}\otimes|m_{k+1}\rangle_b^{(k+1)}\prod_{i=1}^k
\otimes (|n_i\rangle_a^{(i)}|m_i\rangle_b^{(i)}),\cr &&\label{9}
\end{eqnarray}
where we use $\otimes$ and superscript $``(i)"$ to separate and
label the states of distinguishable photons.

Since $|n_i\rangle_a|m_i\rangle_b$ is the same state as $|N_a,
M_b\rangle$ that gives rise to the enhancement factor in
Eq.(\ref{8}), it will contribute a factor of $(n_i+m_i)!/n_i!m_i!$
to the overall enhancement factor, which is then
\begin{eqnarray}
{P_{\{n_i,m_i\}}\over P_{N+M}^{cl}} = \prod_{i=1}^k
{(n_i+m_i)!\over n_i!m_i!}.\label{10}
\end{eqnarray}
Note that since $|n_{k+1}\rangle_a$ and $|m_{k+1}\rangle_b$ are
distinguishable states, they have no contribution to the
enhancement factor.

In Tables I-III, we list the enhancement factors for various
scenarios of the input states $|2_a,2_b\rangle, |3_a,2_b\rangle,
|3_a,3_b\rangle$, respectively. These tables are in contrast with
the visibility tables in Ref.\cite{ou}.

\begin{table}
\caption{\label{tab:table1}Enhancement factor for 2 $a$-photons
and 2 $b$-photons input }
\begin{ruledtabular}
\begin{tabular}{cccc}
2$a$2$b$&2$a$1$b$ + 1$b$
&1$ab$+1$ab$&1$ab$+$a+b$\\
\hline
6 & 3 & 4& 2\\
\end{tabular}
\end{ruledtabular}
\end{table}

\begin{table}
\caption{\label{tab:table2}Enhancement factor for 3 $a$-photons
and 2 $b$-photons input }
\begin{ruledtabular}
\begin{tabular}{cccccccc}
3$a$2$b$&2$a$2$b$&3$a$1$b$&2$a$1$b$ &1$a$2$b$&2$a$1$b$
&$ab$+$a$&$ab$+$a$\\&+$a$&+$b$&+$ab$&+$2a$ &+$a+b$&+$ab$&
+$a+b$\\\hline
10&6&4&6&3&3&4&2\\
\end{tabular}
\end{ruledtabular}
\end{table}

\begin{table*}
\caption{Enhancement factor for 3 $a$-photons and 3 $b$-photons
input}
\begin{ruledtabular}
\begin{tabular}{ccccccccccc}
\noalign{\smallskip} 3$a3b$&3$a2b$&$3a1b$&$2a2b$
&2$a$2$b$&2$a1b$&2$a1b$&2$a$1$b$&$a$$b$$\times$3&$a$$b$$\times$2&$a$$b$+$b$
\\&+$b$&+2$b$&+$a$$b$&+$a$+$b$&+1$a$2$b$&+1$ab+b$&+$a+b+b$&
&+$a$+$b$&+$a$+$a$+$b$\\ \noalign{\smallskip}\hline
\noalign{\smallskip}
 20 & 10 & 4&12&6&9&6&3&8&4&2 \\
\end{tabular}
\end{ruledtabular}
\end{table*}

\section{Summary and Discussion}

In this paper, we discussed a generalized photon bunching effect
which may involve arbitrary number of photons. This bunching
effect is a result of constructive multi-photon interference and
is responsible for stimulated emission of an excited atom.
Furthermore, we find the bunching effect can be used to
characterize temporal distinguishability of photons: various
scenarios of photon temporal distribution give different
enhancement factors.

From Eq.(\ref{10}), we find that the larger the photon number, the
larger the enhancement factor and it is largest for the case of
$N=M$. The largeness of the enhancement factor will make the
measurement process easier. Although the enhancement factor in
Eq.(\ref{10}) does not depend on $T,R$, the actual value of the
probability in Eq.(\ref{6}) does and it is maximum when
$T=N/(N+M)$. From Tables I-III in comparison with the tables in
Ref.\cite{ou}, we find that the values here are more spreading out
and this also makes it easier to tell them apart experimentally.

In this paper, we only considered some extreme cases, i.e., the
photons are either completely indistinguishable, as described in
Eq.(\ref{perm}), or completely distinguishable, as in
Eq.(\ref{orth}). The intermediate case is very complicated and it
is hard to derive the enhancement factor in a closed form. For
some simple cases, however, we are able to do it. For example,
Eq.(4.20) in Ref.\cite{rhe2} provides a formula for the
enhancement factor for the input state of $|2_a, 2_b\rangle$ from
parametric down-conversion. It was found that the enhancement
factor depends on the quantity ${\cal E/A}$, which characterizes
the degree of distinguishability between different pairs of
photons in parametric down-conversion.

\begin{acknowledgments}
This work was supported by the US National Science Foundation
under Grant No. 0427647.
\end{acknowledgments}

\end{document}